\documentclass[fleqn]{revtex4}

\usepackage{amsmath}
\usepackage{graphicx}
\usepackage{multirow}
\newcommand{\bN}{\mbox{\boldmath$\nabla$}}
\newcommand{\br}{\mathbf{r}}

\begin{document}

\title{One-loop vacuum polarization at $m\alpha^7$ and higher orders for three-body molecular systems}

\author{J.-Ph.~Karr}
\author{L.~Hilico}
\affiliation{Laboratoire Kastler Brossel, UPMC-Univ. Paris 6, ENS, CNRS, Coll\`ege de France\\
4 place Jussieu, F-75005 Paris, France}
\affiliation{Universit\'e d'Evry-Val d'Essonne, Boulevard Fran\c cois Mitterrand, F-91000 Evry, France}
\author{Vladimir I. Korobov}
\affiliation{Bogoliubov Laboratory of Theoretical Physics, Joint Institute
for Nuclear Research, Dubna 141980, Russia}

\begin{abstract}
We present calculations of the one-loop vacuum polarization correction (Uehling potential) for the three-body problem in the NRQED formalism. The case of one-electron molecular systems is considered. Numerical results of the vacuum polarization contribution at $m\alpha^7$ and higher orders for the fundamental transitions $(v=0,L=0)\to(v'=1,L'=0)$ in the H$_2^+$ and HD$^+$ molecular ions are presented and compared with calculations performed in the adiabatic approximation. The residual uncertainty from this contribution on the transition frequencies is shown to be of a few tens of~Hz.
\end{abstract}

\maketitle

\section*{Introduction}

The hydrogen molecular ions $H_2^+$ and HD$^+$ have great potentiality for improving the determination of fundamental constants such as the proton-to-electron mass ratio~\cite{Wing76,Gremaud98,Karr16}. Ro-vibrational transition frequencies have been measured at the few-ppb level~\cite{Koelemeij07,Bressel12,Biesheuvel16}, and ongoing efforts towards improved accuracies using two-photon transitions~\cite{Karr16,Karr12,Tran13} or one-photon transitions in the Lamb-Dicke regime~\cite{Schiller14,Karr14b} motivate the development of precise theoretical predictions.

In Refs.~\cite{Korobov14PRA,Korobov14} a complete set of $m\alpha^7$-order contributions has been evaluated for the fundamental transitions of the hydrogen molecular ions H$_2^+$ and HD$^+$. All calculations at this order were performed in the nonrecoil limit, by evaluating the one-electron QED corrections in the two-center approximation. Only one term, the Uehling potential vacuum polarization contribution~\cite{Uehling}, which had been estimated with a lower level of accuracy, was later calculated in the framework of the two-center approximation in~\cite{Karr14}.

In a slightly different context, namely the hyperfine structure of H$_2^+$, it was recently shown~\cite{Korobov16} that in evaluating a second-order perturbation term within the $m\alpha^6(m/M)$-order relativistic correction it is essential to take the vibrational motion of nuclei into account. Such vibrational contributions also arise in the spin-independent corrections, and have to be consistently included in the previously evaluated~\cite{Korobov08,Korobov14PRA,Korobov14,Karr14} $m\alpha^6$ and $m\alpha^7$-order corrections~\cite{Korobov16b}.

The $m\alpha^7$-order Uehling contribution~\cite{Karr14} is one such case. In the present work we first revisit the evaluation of this term within the adiabatic approximation by including the previously omitted vibrational contribution. Then we go one step further and evaluate it in a full three-body approach, exploiting the fact that the matrix elements of the Uehling potential in a basis of explicitly correlated exponential functions are known in analytical form~\cite{Karr13}. Comparison of results obtained with these two approaches provide a useful cross-check and give interesting insight on the precision of the adiabatic approximation for evaluating QED corrections in molecular systems.

\section{Uehling correction terms at $m\alpha^7$ and higher orders}

\subsection{General expressions} \label{general}

We use atomic units throughout. The system under consideration is composed of three particles with masses $m_i$ and charges $Z_i$ ($i=1,2,3$). We specifically consider a molecular or molecule-like system and assume that the lightest particle -i.e. an electron in the practical cases considered here- is numbered 3 (thus $m_1, m_2 \gg m_3 = m_e$, and $Z_3 = -1$). The relative positions of particles 1-3 and 2-3 (electron-nucleus) are respectively denoted by $\br_1$ and $\br_2$, and the relative position of particles 1-2 (internuclear) by $\br_{12}$. Whenever the adiabatic approximation is used, we will set $r_{12} = R$.

The correction terms to be considered are the same as those studied in the two-center approximation in~\cite{Karr14}, but we will use slightly different notations to make the comparison between adiabatic and full three-body results more transparent. All terms involve the Uehling potential interaction between the electron and nuclei:
\begin{equation}
U_{vp}(\mathbf{r}) = U_{vp}(r_1) + U_{vp}(r_2)
\end{equation}
where $U_{vp}$ is given by \cite{ItzZuber}:
\begin{equation}
U_{vp}(r_i) = -\frac{2}{3}\frac{Z_i \alpha}{\pi r_i}
   \int_1^\infty dt\> e^{-\frac{2r_i}{\alpha}\,t}
   \left(
      \frac{1}{t^2}+\frac{1}{2t^4}
   \right)
   \left(t^2-1\right)^{1/2}.
\end{equation}
We neglect all corrections originating from the internuclear Uehling interaction, as was done in calculation of lower-order terms~\cite{Korobov06}.

The first correction term comes from the first-order correction with the nonrelativistic wave function $\psi_0$,
\begin{equation}
\Delta E_a = \left\langle \psi_0 | U_{vp} | \psi_0 \right\rangle \label{ea}.
\end{equation}

The second contribution comes from the relativistic correction to the wave function. It takes the form of a second-order contribution with the Breit-Pauli Hamiltonian $H_B$ as the perturbation:
\begin{equation}
\Delta E_b = 2 \left\langle \psi_0 | H_B Q (E_0 - H_0)^{-1} Q U_{vp} | \psi_0 \right\rangle \label{eb}.
\end{equation}
Here, $Q = I - | \psi_0 \rangle \langle \psi_0 | $ is a projection operator, $H_0$ and $E_0$ the nonrelativistic Hamiltonian and energy, and $H_B$ is the spin-independent relativistic correction to the electron
\begin{equation}
H_B = -\frac{\mathbf{p_e}^4}{8 m_e^3} + \frac{\pi}{2 m_e^2} \left[ Z_1 \delta(\mathbf{r_1}) + Z_2 \delta(\mathbf{r_2}) \right] .
\end{equation}
For a full three-body treatment, one should take as $H_B$ the full three-body Breit-Pauli Hamiltonian. However, our goal is to analyze the accuracy of the two-center approximation, which is why we include the exact same relativistic corrections in both approaches. The neglected radiative-recoil terms of orders $m\alpha^7 (m/M)^n, \; n=1,2...$ ($m \equiv m_e$, $M \equiv m_1,m_2$) are much smaller and irrelevant at the current level of theoretical accuracy.

The last contribution is the vertex function modification (Darwin term) at $m\alpha^7$ order (see Fig.~3 in \cite{Kinoshita96}):
\begin{equation}
\Delta E_c = \langle \psi_0 | H^{(7)}_{vp} | \psi_0 \rangle, \label{ec}
\end{equation}
\begin{equation}
H^{(7)}_{vp} = \frac{1}{8 m_e^2} ( \Delta_{\br_1} U_{vp}(r_1) + \Delta_{\br_2} U_{vp}(r_2) ). \label{heff}
\end{equation}
In a full three-body treatment, additional radiative-recoil terms with $m_1^2$ and $m_2^2$ at the denominator instead of $m_e^2$ should be included. Similarly to the $\Delta E_b$ contribution discussed above, we neglect these terms here. Finally, we also neglect the transverse photon exchange and spin-orbit terms~\cite{Pachucki96} which produce corrections of order $m\alpha^7 (m/M)$.

The total Uehling energy correction is
\begin{equation}
\Delta E_U = \Delta E_a + \Delta E_b + \Delta E_c. \label{evp}
\end{equation}
Each of the three contributions contains lower-order terms ($m \alpha^5$, $m \alpha^6$) which should be subtracted in order to get the desired contribution ($m\alpha^7$ and above). This subtraction procedure will be explained in the next paragraphs, first in the adiabatic approximation and then for the three-body case.

\subsection{Adiabatic approximation}

In this approach, $\psi_0$ is an adiabatic wave function given by
\begin{equation}
\psi_0 = \phi_{el}(\mathbf{r};R) \chi(R) \label{bo}
\end{equation}
where $\phi_{el}$ and $\chi$ are respectively the electronic and nuclear wave functions. The Hamiltonian $H_0$ appearing in Eq.~(\ref{eb}) is an adiabatic Hamiltonian, and $E_0$ the adiabatic energy (see e.g.~\cite{Wolniewicz80} for definitions).

Within the adiabatic approximation, the second-order perturbation term $\Delta E_b$ can be separated into electronic and vibrational contributions~\cite{Korobov16,Korobov16b}:
\begin{eqnarray}
\Delta E_b &=& \left\langle \chi \left| \; \mathcal{E}^{(el)}_b(R) \right| \chi \right\rangle + E^{(vb)}_b, \label{ebsepa} \\
\mathcal{E}^{(el)}_b(R) &=& 2 \left\langle \phi_{el} | H_B Q_{el} (E_{el} - H_{el})^{-1} Q_{el} U_{vp} | \phi_{el} \right\rangle \label{ebelec} \\
E^{(vb)}_b &=& 2 \left\langle \chi | \mathcal{E}_B (R) Q_{vb} (E_{vb} - H_{vb})^{-1} Q_{vb} \mathcal{E}_{vp} (R) | \chi \right\rangle \label{ebvibr}
\end{eqnarray}
$Q_{el} = I - | \phi_{el} \rangle \langle \phi_{el} |$ and $Q_{vb} = I - | \chi \rangle \langle \chi |$ are projection operators, and $H_{el}$, $E_{el}$ (resp. $H_{vb}$, $E_{vb}$) the electronic (resp. vibrational) Hamiltonian and energy. Finally $\mathcal{E}_B (R) = \left\langle \phi_{el} | H_B | \phi_{el} \right\rangle$, and $\mathcal{E}_{vp} (R) = \left\langle \phi_{el} | U_{vp} | \phi_{el} \right\rangle$. Only the first term of Eq.~(\ref{ebsepa}) was calculated in Ref.~\cite{Karr14}, while the vibrational contribution was omitted.

The expansion in powers of $\alpha$ of each term in Eq.~(\ref{evp}) was studied in~\cite{Karr14}. We reproduce the results here for convenience:
\begin{eqnarray}
\Delta E_a &=& - \frac{4\alpha^3}{15} \left\langle \psi_0 | Z_1 \delta(\mathbf{r_1}) + Z_2 \delta(\mathbf{r_2}) | \psi_0 \right\rangle
   + \frac{5\alpha^4}{48} \pi \left\langle \psi_0 | Z_1^2 \delta(\mathbf{r_1}) + Z_2^2 \delta(\mathbf{r_2}) | \psi_0 \right\rangle + \ldots \label{expa} \\
\Delta E_b &=& - \frac{3\alpha^4}{16} \pi \left\langle \psi_0 | Z_1^2 \delta(\mathbf{r_1}) + Z_2^2 \delta(\mathbf{r_2}) | \psi_0 \right\rangle + \frac{4 \alpha^5}{15} \ln \alpha \left\langle \psi_0 | Z_1^3 \delta(\mathbf{r_1}) + Z_2^3 \delta(\mathbf{r_2}) | \psi_0 \right\rangle \ldots \\
\Delta E_c &=& + \frac{3\alpha^4}{16} \pi \left\langle \psi_0 | Z_1^2 \delta(\mathbf{r_1}) + Z_2^2 \delta(\mathbf{r_2}) | \psi_0 \right\rangle + \ldots \label{expc}
\end{eqnarray}
The first two terms of $\Delta E_a$ are the leading terms of the Uehling correction, which were already included in earlier calculations~\cite{Korobov06}. Indeed, the $m\alpha^6$-order terms appearing in $\Delta E_b$ and $\Delta E_c$ cancel each other. Note that this exact cancellation no longer occurs in the three-body approach, as will be seen below.

The sought corrections of order $m\alpha^7$ and above (excluding the logarithmic contribution in $\Delta E_b$, which was already considered in~\cite{Korobov14PRA,Korobov14}), are thus given by the following subtractions:
\begin{eqnarray}
\Delta E_U^{(7+)} &=& \Delta E_a^{(7+)} + \Delta E_b^{(7+)} + \Delta E_c^{(7+)}, \\
\Delta E_a^{(7+)} &=& \Delta E_a + \frac{4\alpha^3}{15} \left\langle \psi_0 | Z_1 \delta(\mathbf{r_1}) + Z_2 \delta(\mathbf{r_2}) | \psi_0 \right\rangle \label{e7a-ad}
   - \frac{5\alpha^4}{48} \pi \left\langle \psi_0 | Z_1^2 \delta(\mathbf{r_1}) + Z_2^2 \delta(\mathbf{r_2}) | \psi_0 \right\rangle \\
\Delta E_b^{(7+)} &=& \Delta E_b + \frac{3\alpha^4}{16} \pi \left\langle \psi_0 | Z_1^2 \delta(\mathbf{r_1}) + Z_2^2 \delta(\mathbf{r_2}) | \psi_0 \right\rangle - \frac{4 \alpha^5}{15} \ln \alpha \left\langle \psi_0 | Z_1^3 \delta(\mathbf{r_1}) + Z_2^3 \delta(\mathbf{r_2}) | \psi_0 \right\rangle\\
\Delta E_c^{(7+)} &=& \Delta E_c - \frac{3\alpha^4}{16} \pi \left\langle \psi_0 | Z_1^2 \delta(\mathbf{r_1}) + Z_2^2 \delta(\mathbf{r_2}) | \psi_0 \right\rangle \label{e7c-ad}
\end{eqnarray}
Note that the definitions of $\Delta E_b^{(7+)}$ and $\Delta E_c^{(7+)}$ differ from those of Ref.~\cite{Karr14}. Finally, since the leading-order terms in the expansion $\Delta E_b$ belong to the electronic contribution, we define the higher-order electronic contribution as
\begin{equation}
\Delta E_b^{(7+)(el)} = \left\langle \chi \left| \; \mathcal{E}^{(el)}_b(R) \right| \chi \right\rangle + \frac{3\alpha^4}{16} \pi \left\langle \psi_0 | Z_1^2 \delta(\mathbf{r_1}) + Z_2^2 \delta(\mathbf{r_2}) | \psi_0 \right\rangle - \frac{4 \alpha^5}{15} \ln \alpha \left\langle \psi_0 | Z_1^3 \delta(\mathbf{r_1}) + Z_2^3 \delta(\mathbf{r_2}) | \psi_0 \right\rangle.
\end{equation}

\subsection{Three-body formalism}

In this approach, $H_0$ is the exact nonrelativistic Hamiltonian of the three-body system, i.e.
\begin{equation}
H_0 = -\frac{1}{2\mu_{13}}\bN^2_{\br_1}-\frac{1}{2\mu_{23}}\bN^2_{\br_2} - \bN_{\br_1}\bN_{\br_2} - \frac{Z_1}{r_1} - \frac{Z_2}{r_2} + \frac{Z_1Z_2}{r_{12}},
\end{equation}
where $\mu_{ij} = m_i/(m_i+m_j)$, and $\psi_0$ is one of its eigenstates. In this case, the expansion of Uehling correction terms in powers of $\alpha$ is modified with respect to Eqs.~(\ref{expa}-\ref{expc}):
\begin{eqnarray}
\Delta E_a &=& - \frac{4\alpha^3}{15} \left\langle \psi_0 | Z_1 \delta(\mathbf{r_1}) + Z_2 \delta(\mathbf{r_2}) | \psi_0 \right\rangle \label{eaexp}
   + \frac{5\alpha^4}{48} \pi \left\langle \psi_0 | \mu_{13} Z_1^2 \delta(\mathbf{r_1}) + \mu_{23} Z_2^2 \delta(\mathbf{r_2}) | \psi_0 \right\rangle + \ldots \\
\Delta E_b &=& - \frac{3\alpha^4}{16} \pi \left\langle \psi_0 | a_{13} Z_1^2 \delta(\mathbf{r_1}) + a_{23} Z_2^2 \delta(\mathbf{r_2}) | \psi_0 \right\rangle + \frac{4 \alpha^5}{15} \ln \alpha \left\langle \psi_0 | \mu_{13} a_{13} Z_1^3 \delta(\mathbf{r_1}) + \mu_{23} a_{23} Z_2^3 \delta(\mathbf{r_2}) | \psi_0 \right\rangle \ldots \label{ebexp} \\
\Delta E_c &=& + \frac{3\alpha^4}{16} \pi \left\langle \psi_0 | \mu_{13} Z_1^2 \delta(\mathbf{r_1}) + \mu_{23} Z_2^2 \delta(\mathbf{r_2}) | \psi_0 \right\rangle + \ldots \label{ecexp}
\end{eqnarray}
where $a_{ij} = \mu_{ij} (2 \mu_{ij} - 1)$. These modifications can be understood as follows. Both for $\Delta E_a$ and $\Delta E_c$, the successive terms of the $\alpha$-expansion are proportional to the successive derivatives of the squared wave function at the electron-nucleus coalescence points. In the second term of $\Delta E_a$ and in the first term of $\Delta E_c$, both of which involve the first derivative, the appearance of the additional factors $\mu_{13}$, $\mu_{23}$ comes from Kato's cusp condition~\cite{Kato57} in the case of a finite nuclear mass:
\begin{equation}
\left. \frac{\partial \psi_0}{\partial r_i} \right|_{r_i = 0} = - \mu_{i3} Z_i \psi_0 (r_i = 0), \hspace{2mm} i=1,2.
\end{equation}
As for $\Delta E_b$, Eq.~(\ref{ebexp}) can be understood by writing this term in the following equivalent form:
\begin{equation}
\Delta E_b = 2 \left\langle \psi_B | U_{vp} | \psi_0 \right\rangle, \label{eb-with-psib}
\end{equation}
where $\psi_B$ is the first-order correction to the wave function induced by the relativistic correction $H_B$:
\begin{equation}
(E_0 - H_0 ) \psi_B = \left( H_B - \langle H_B \rangle \right) \psi_0 . \label{psib}
\end{equation}
It was shown in~\cite{Karr14} that the $m\alpha^6$ and $m\alpha^7 \ln \alpha$ terms of $\Delta E_b$ respectively come from the $1/r_i$ and $\ln r_i$ singularities of $\psi_B$. The analysis of Eq.~(\ref{psib}) in the limit $r_i \to 0$ reveals that the singular parts of $\psi_B$ write~\cite{Korobov09}
\begin{equation}
\psi_B^{\rm sing} = (U_1 - \langle U_1 \rangle) \Psi_0 \; , \qquad U_1 = \frac{a_{13} Z_1}{4 r_1} + \frac{a_{23} Z_2}{4 r_2},
\end{equation}
\begin{equation}
\psi_B^{\rm log} = (U_2 - \langle U_2 \rangle) \Psi_0 \; , \qquad U_2 = -\frac{\mu_{13} a_{13} Z_1^2}{2} \ln r_1 -\frac{\mu_{23} a_{23} Z_2^2}{2} \ln{r_2},
\end{equation}
which explains the factors appearing in the first terms of the $\alpha$-expansion.

One can observe that the $m\alpha^6$-order terms in $\Delta E_b$ and $\Delta E_c$ no longer cancel, but their sum produces recoil terms. Overall, the correction $\Delta E_U$ contains a set of recoil corrections at orders $m\alpha^6 (m/M)^n$. Note that the latter do not add up to yield the known result for the $m\alpha^6$-order term including recoil effects~\cite{Korobov06,Eides,Codata10}, because some recoil contributions are missing due to the neglected terms in $H_{vp}^{(7)}$ (see Sec.~\ref{general}). This is of no consequence here, since $m\alpha^6$-order terms are subtracted in order to  focus on corrections of order $m\alpha^7$ and above.

Similarly, $\Delta E_U^{(7+)}$ contains an incomplete set of recoil corrections at orders $m\alpha^7 (m/M)^n$, therefore the results obtained in the three-body framework are expected to be accurate to $\mathcal{O}(m/M)$, just as within the adiabatic approximation.

The expansions~(\ref{eaexp}-\ref{ecexp}) lead to the following definitions for the corrections of order $m\alpha^7$ and above:
\begin{eqnarray}
\Delta E_a^{(7+)}\!&=&\!\Delta E_a\! +\! \frac{4\alpha^3}{15} \left\langle \psi_0 | Z_1 \delta(\mathbf{r_1}) \!+\! Z_2 \delta(\mathbf{r_2}) | \psi_0 \right\rangle \label{e7a-3body}
   \!-\! \frac{5\alpha^4}{48} \pi \left\langle \psi_0 | \mu_{13} Z_1^2 \delta(\mathbf{r_1}) \!+\! \mu_{23} Z_2^2 \delta(\mathbf{r_2}) | \psi_0 \right\rangle \\
\Delta E_b^{(7+)}\! &=&\! \Delta E_b \!+\! \frac{3\alpha^4}{16} \pi \left\langle \psi_0 | a_{13} Z_1^2 \delta(\mathbf{r_1}) \!+\! a_{23} Z_2^2 \delta(\mathbf{r_2}) | \psi_0 \right\rangle \!-\! \frac{4 \alpha^5}{15} \ln \alpha \left\langle \psi_0 | \mu_{13} a_{13} Z_1^3 \delta(\mathbf{r_1}) \!+\! \mu_{23} a_{23} Z_2^3 \delta(\mathbf{r_2}) | \psi_0 \right\rangle \\
\Delta E_c^{(7+)}\! &=& \!\Delta E_c \!-\! \frac{3\alpha^4}{16} \pi \left\langle \psi_0 | \mu_{13} Z_1^2 \delta(\mathbf{r_1}) \!+\! \mu_{23} Z_2^2 \delta(\mathbf{r_2}) | \psi_0 \right\rangle. \label{e7c-3body}
\end{eqnarray}

\section{Numerical calculations and results}

In this Section we calculate and compare the Uehling corrections obtained within the adiabatic (Eqs.~(\ref{e7a-ad}-\ref{e7c-ad})) and three-body (Eqs.~(\ref{e7a-3body}-\ref{e7c-3body})) approaches.

\subsection{Adiabatic approximation}

For the adiabatic case all the corrections terms, with the exception of the vibrational contribution in $\Delta E_b$ (Eq.~(\ref{ebvibr})), have been evaluated in our previous work~\cite{Karr14} and more details may be found in that reference.

Here we only recall the main features of our approach. In the spirit of the adiabatic approximation, in a first step we calculate the electronic curves corresponding to the correction terms: $\mathcal{E}_{vp} (R)$, (defined after Eq.~(\ref{ebvibr})), $\mathcal{E}_b^{(el)} (R)$ (Eq.~(\ref{ebelec})), and
\begin{eqnarray}
\mathcal{E}_c (R) &=& \langle \phi_{el} | H^{(7)}_{vp} | \phi_{el} \rangle, \label{ecelec}
\end{eqnarray}
as well as $\mathcal{E}_B (R)$,  which is required for the evaluation of $\Delta E_b^{(vb)}$ (see Eq.~(\ref{ebvibr})).

We use the following variational expansion for the electronic wave function of a $\sigma$ state:
\begin{equation}
\phi_{el} (\mathbf{r}) = \sum_{i=1}^{\infty} C_i e^{-a_i r_1 - b_i r_2}, \label{basis-set}
\end{equation}
which is symmetrized if $Z_1 = Z_2$:
\begin{equation}
\phi_{el} (\mathbf{r}) = \sum_{i=1}^{\infty} C_i \left( e^{-a_i r_1 - b_i r_2} \pm e^{-b_i r_1 - a_i r_2} \right).
\end{equation}
The real exponents $a_i$ and $b_i$ are generated in a quasi-random manner in optimized intervals.

We now describe the improvements we have implemented with respect to the calculations presented in Ref.~\cite{Karr14}. First of all, we discovered that the transformation of the $\Delta E_c^{(7+)}$ term using integration by parts (Eq.~(22) of~\cite{Karr14}) is not valid for a two-center system (although it is valid for a hydrogenlike atom) leading to a numerical error of a few~kHz. We have thus recalculated $\mathcal{E}_c(R)$ directly from Eq.(~\ref{ecelec}).

For the electronic contribution to $\Delta E_b$ (Eq.~(\ref{ebelec})) we use the equivalent form
\begin{equation}
\mathcal{E}^{(el)}_b(R) = 2 \left\langle \phi_B | U_{vp} | \phi_{el} \right\rangle,
\end{equation}
where $\phi_B$ is the first-order correction to the electronic wave function induced by the relativistic correction $H_B$:
\begin{equation}
(E_{el} - H_{el}) \phi_B = \left( H_B - \langle H_B \rangle \right) \phi_{el}. \label{psib-elec}
\end{equation}
Trying to calculate $\phi_B$ directly by solving the linear problem~(\ref{psib-elec}) would lead to numerical problems, because $\phi_B$ contains singular terms (in $1/r_i$ and $\ln r_i$, $i=1,2$) which are not well represented in the regular basis set~(\ref{basis-set}). We thus separate the singular terms in $\phi_B$ following the approach described in~\cite{Korobov13}:
\begin{equation}
\phi_B (r_1,r_2) = \left( \frac{Z_1}{4 r_1} + \frac{Z_2}{4 r_2} - \frac{Z_1^2}{2} \ln r_1 - \frac{Z_2^2}{2} \ln r_2 \right) \phi_{el} + \tilde{\phi}_B (r_1,r_2),
\end{equation}
where $\tilde{\phi}_B (r_1,r_2)$ is a regular function which is obtained numerically by solving the linear problem
\begin{eqnarray}
(E_{el} - H_{el}) \tilde{\phi}_B &=& \left( H_B - \langle H_B \rangle \right) \phi_{el} + \left[ H , \left( \frac{Z_1}{4 r_1} + \frac{Z_2}{4 r_2} - \frac{Z_1^2}{2} \ln r_1 - \frac{Z_2^2}{2} \ln r_2 \right) \right] \phi_{el} \nonumber \\
&=& \left( H_B - \langle H_B \rangle \right) \phi_{el} + \sum_{i=1,2} \left( \frac{Z_i \pi \delta(\mathbf{r_i})}{2} + \frac{Z_i^2}{4 r_i^2} + \frac{Z_i \mathbf{r_i}}{4 r_i^3} \boldsymbol{\nabla} + \frac{Z_i^2 \mathbf{r_i}}{2 r_i^2} \boldsymbol{\nabla} \right) \phi_{el}.
\end{eqnarray}
Finally, one obtains
\begin{eqnarray}
\mathcal{E}^{(el)}_b(R) &=& 2 \left\langle \left( \frac{Z_1}{4 r_1} + \frac{Z_2}{4 r_2} - \frac{Z_1^2}{2} \ln r_1 - \frac{Z_2^2}{2} \ln r_2 \right) U_{vp} \right\rangle
 - 2 \left\langle \left( \frac{Z_1}{4 r_1} + \frac{Z_2}{4 r_2} - \frac{Z_1^2}{2} \ln r_1 - \frac{Z_2^2}{2} \ln r_2 \right) \right\rangle \left\langle U_{vp} \right\rangle \nonumber \\
 && + 2 \left\langle \tilde{\phi}_B | U_{vp} | \phi_{el} \right\rangle. \label{eb-final}
\end{eqnarray}

The terms involving the Uehling potential (i.e. $\mathcal{E}_{vp}(R)$, $\mathcal{E}_b^{(el)}(R)$ and $\mathcal{E}_c (R)$) cannot be calculated exactly since its matrix elements in the exponential basis set~(\ref{basis-set}) are not known in analytical form. We calculated them by two different methods: (i) by numerical integration as was done in~\cite{Karr14}, using an approximate form of the Uehling potential presented in~\cite{Fullerton76} which is accurate to at least nine digits, and (ii) by expanding the matrix elements in powers of $\alpha$, which allows for much quicker calculations. The expansions of all the required matrix elements are given in the Appendix. We included all terms up to the $m\alpha^8$ order in our calculation, and found excellent agreement with the method (i) (see the Appendix for a numerical example), thus removing any doubt that may arise on the accuracy of the numerical integration.

Finally, in a second step the electronic curves are averaged over the vibrational wavefunction $\chi(R)$ which is obtained by numerical resolution of the nuclear Schr\"odinger equation. The vibrational contribution $E_b^{(vb)}$ (Eq.~(\ref{ebvibr})) is obtained using the first-order relativistic correction $\chi_B$ to the nuclear wave function:
\begin{equation}
E^{(vb)}_b = 2 \left\langle \chi_B | \mathcal{E}_{vp} (R) | \chi \right\rangle,
\end{equation}
where $\chi_B$ is calculated by solving the linear problem
\begin{equation}
(E_{vb} - H_{vb}) \chi_B = \mathcal{E}_B (R) \chi.
\end{equation}

\subsection{Three-body formalism}

For the three-body case we used a variational ''exponential'' expansion of the three-body wavefunction in the form~\cite{Korobov00}
\begin{equation}\label{expansion}
\Psi \left( \br_1,\br_2,\br_{12} \right) = \sum_{n=1}^{N}
\Bigl\{
   U_i{\rm{Re}}[e^{-\alpha_ir_1-\beta_ir_2-\gamma_ir_{12}}]
   +W_i{\rm{Im}}[e^{-\alpha_ir_1-\beta_ir_2-\gamma_ir_{12}}]
\Bigr\}\,
  {\mathcal{Y}}^{l_1,l_2}_{LM}(\hat{\bf{r}}_1,\hat{\bf{r}}_2)\,,
\end{equation}
where $\mathcal{Y}^{l_1l_2}_{LM}(\hat\br_1,\hat\br_2)$ are bipolar spherical harmonics \cite{Varshalovich}. Parameters $\alpha_i$, $\beta_i$, $\gamma_i$ are complex exponents satisfying the relations ${\rm Re}(\alpha_i + \beta_i) > 0$, ${\rm Re}(\alpha_i + \gamma_i) > 0$, and ${\rm Re}(\beta_i + \gamma_i) > 0$, generated in a pseudorandom way \cite{Smith77} in several intervals; the variational parameters are the bounds of these intervals.

Here we consider only rotationless ($L = 0$) states. For these states the matrix elements of the Uehling potential $U_{vp}(r_i)$ required for calculation of $\Delta E_a$ were obtained in~\cite{Karr13}, and those of $\Delta_{\br_1} U_{vp}(r_1)$ required for $\Delta E_c$ are given in the Appendix.

The precise calculation of the second-order perturbation term $\Delta E_b$ is more challenging, because it involves solving the linear problem~(\ref{psib}). Similarly to what was done for the two-center problem, we separate the singular part of $\psi_B$ in order to alleviate the numerical difficulties. We introduce a less singular function $\tilde{\psi}_B$ defined by
\begin{equation}
\psi_B = \psi_B^{\rm sing} + \tilde{\psi}_B,
\end{equation}
and $\tilde{\psi}_B$ can be obtained by solving the equation
\begin{equation}
(E_0 - H_0) \tilde{\Psi}_B = \left( H_B - \langle H_B \rangle \right) \Psi_0 + [ H_0, U_1 ] \Psi_0,
\end{equation}
Straightforward algebraic manipulation leads to
\begin{equation}
[ H_0, U_1 ] =
  \frac{Z_1 a_{13}}{4} \left[ \frac{1}{\mu_{13}}
     \left\{ 2\pi\delta(\br_1) + \frac{\br_1 \cdot \bN_{\br_1}}{r_1^3} \right\}
  + \frac{\br_1 \cdot \bN_{\br_2}}{r_1^3} \right]
  +\frac{Z_2 a_{23}}{4} \left[ \frac{1}{\mu_{23}}
     \left\{ 2\pi\delta(\br_2) + \frac{\br_2 \cdot \bN_{\br_2}}{r_2^3} \right\}
  + \frac{\br_2 \cdot \bN_{\br_1}}{r_2^3} \right].
\end{equation}
The final expression of the second-order perturbation term is
\begin{eqnarray}
\Delta E_b &=& 2 \left\langle \psi_0 | U_1 U_{vp} | \psi_0 \right\rangle - 2 \langle U_1 \rangle \langle \psi_0 | U_{vp} | \psi_0 \rangle + 2 \left\langle \tilde{\psi}_B | U_{vp} | \psi_0 \label{eb3b-final} \right\rangle.
\end{eqnarray}
The calculation of the first term requires the matrix elements of $U_{vp}(r_i)/r_i$, which are given in the Appendix, and crossed terms of the type $U_{vp}(r_i)/r_j$ whose matrix elements are easily obtained from the generating integral given in~\cite{Karr13}. It should be noted that in contradistinction with the two-center case, we have separated the $1/r$ singularities of~(\ref{psib}) but not the logarithmic ones. Due to this the convergence of $\Delta E_b$ is much slower. The separation of the logarithmic singularity would require the derivation of three-body matrix elements involving logarithms of inter-particle distances.

\subsection{Results and discussion}

\begin{table}[t]
\begin{center}
\begin{tabular}{|@{\hspace{2mm}}c@{\hspace{2mm}}|@{\hspace{2mm}}r@{\hspace{2mm}}|@{\hspace{2mm}}r@{\hspace{2mm}}r@{\hspace{2mm}}r@{\hspace{2mm}}|@{\hspace{2mm}}r@{\hspace{2mm}}|@{\hspace{2mm}}r@{\hspace{2mm}}|}
\hline\hline
                                     & $\Delta E_a^{(7+)}$ & electr.  & vibr.   & $\Delta E_b^{(7+)}$ & $\Delta E_c^{(7+)}$ & $\Delta E_U^{(7+)}$ \\
\hline
\multirow{2}{*}{H$_2^+$ $(v=0,L=0)$} &      $-$20.11       &  15.26   & $-$4.61 &    10.64            &    $-$23.49         &      $-$32.95     \\
                                     &      $-$20.06       &    -     &    -    &    10.64            &    $-$23.43         &      $-$32.86     \\
\hline
\multirow{2}{*}{H$_2^+$ $(v=1,L=0)$} &     $-$19.54        &  14.44   & $-$4.18 &    10.26            &    $-$22.83         &      $-$32.10      \\
                                     &     $-$19.49        &    -     &    -    &    10.27            &    $-$22.77         &      $-$32.00      \\
\hline
\multirow{2}{*}{$v=0 \to 1$ transition} &  0.57            &  -0.81   &  0.43   &  $-$0.38            &       0.66          &         0.85        \\
                                     &     0.57            &    -     &    -    &  $-$0.38            &       0.65          &         0.85        \\

\hline\hline
\end{tabular}
\end{center}
\caption{Uehling corrections at order $m\alpha^7$ and above, in kHz, for the two lowest vibrational states of H$_2^+$ and for the fundamental vibrational transition. For each contribution, the value obtained in the adiabatic approximation is given in the first line, and that obtained within the three-body formalism in the second line. Note that the vibrational part of $\Delta E_b^{(7+)}$ was not included in previous calculations~\cite{Karr14}.\label{results-h2p}}
\end{table}

\begin{table}[t]
\begin{center}
\begin{tabular}{|@{\hspace{2mm}}c@{\hspace{2mm}}|@{\hspace{2mm}}r@{\hspace{2mm}}|@{\hspace{2mm}}r@{\hspace{2mm}}r@{\hspace{2mm}}r@{\hspace{2mm}}|@{\hspace{2mm}}r@{\hspace{2mm}}|@{\hspace{2mm}}r@{\hspace{2mm}}|}
\hline\hline
                                     & $\Delta E_a^{(7+)}$ & electr.  & vibr.   & $\Delta E_b^{(7+)}$ & $\Delta E_c^{(7+)}$ & $\Delta E_U^{(7+)}$ \\
\hline
\multirow{2}{*}{HD$^+$ $(v=0,L=0)$}  &      $-$20.15       &  15.31   & $-$4.65 &    10.67            &    $-$23.54         &      $-$33.02       \\
                                     &      $-$20.11       &    -     &    -    &    10.67            &    $-$23.50         &      $-$32.94       \\
\hline
\multirow{2}{*}{HD$^+$ $(v=1,L=0)$}  &      $-$19.65       &  14.60   & $-$4.27 &    10.34            &    $-$22.96         &      $-$32.27       \\
                                     &      $-$19.62       &    -     &    -    &    10.34            &    $-$22.92         &      $-$32.19       \\
\hline
\multirow{2}{*}{$v=0 \to 1$ transition} &  0.48            &  -0.71   &  0.38   &  $-$0.33            &       0.56          &        0.75         \\
                                     &     0.50            &    -     &    -    &  $-$0.33            &       0.58          &        0.74         \\
\hline\hline
\end{tabular}
\end{center}
\caption{Same as Table \ref{results-h2p}, but for the HD$^+$ molecular ion.\label{results-hdp}}
\end{table}

In order to obtain good convergence of the three-body results, basis sets of $N = 2000$ vectors were used to represent $\psi_0$. For the numerical evaluation of the second-order term (last term of Eq.~(\ref{eb3b-final})) we use ten basis sets, where the first two approximate the regular part of the intermediate solution and the remaining eight sets with growing exponents are introduced to reproduce behavior of the type $\ln{r_1}$ (or $\ln{r_2}$) at small values of $r_1$ ($r_2$). The total size of the basis used for intermediate states is $N = 5900$.

The results for the first vibrational levels of H$_2^+$ and HD$^+$ are presented in Tables~\ref{results-h2p} and \ref{results-hdp}. The relative difference between adiabatic and three-body approaches (2-3$\times$10$^{-3}$) matches the expected order of magnitude $\mathcal{O}(m/M)$ that corresponds to the presence of recoil contributions in the three-body correction. The difference between adiabatic and three-body results also gives an order of magnitude of the residual uncertainty due to unevaluated recoil corrections, i.e. a few tens of~Hz on the transition frequencies. This uncertainty may be reduced further in the future by including all recoil corrections within the three-body approach.

It can also be observed that the inclusion of the vibrational part in the second-order perturbation term $\Delta E_b$ is essential to get satisfactory agreement. This is even more true in the case of vibrational transition frequencies, where this term contributes to about 50\% while representing only 13-14\% of the correction to individual state energies, due to its much stronger dependence on the vibrational state. The corresponding correction to the fundamental vibrational transition amounts to about 400~Hz, which is significant at the current level of theoretical accuracy. It is thus essential to include all terms of similar nature arising in other $m\alpha^7$-order corrections~\cite{Korobov16b}.

\section*{Acknowledgements}

This work was supported by Ecole Normale Supérieure, which is gratefully acknowledged. J.-Ph. Karr acknowledges support as a fellow of the Institut Universitaire de France. V.I.K. acknowledges support of the Russian Foundation for Basic Research under Grant No.~15-02-01906-a.

\section*{Appendix}

\subsection*{Two-center problem: expansion of matrix elements for $\sigma$ electronic states}

In what follows, the notation $\left\langle i | A | j \right\rangle$ stands for the matrix element of the operator $A$ between the basis functions $e^{-a_i r_1 - b_i r_2}$ and $e^{-a_j r_1 - b_j r_2}$. We set $a = a_i + a_j$ and $b = b_i + b_j$. \vspace{3mm}

{\bf 1.} $\mathcal{E}_{vp} (R)$: Uehling potential expectation value
\[
\begin{array}{@{}l}\displaystyle
\left\langle i \bigl| U_{vp}(r_1) \bigr| j \right\rangle =
   -\frac{4}{15}Z_1\alpha^3 e^{-b R}
   \left[
      1-\frac{25\pi\alpha}{128}\,a
      +\frac{3\alpha^2}{28}\left(3a^2\!+\!b^2\!-\!\frac{2b}{R}\right)
      -\frac{105\pi\alpha^3}{2048}\,a\left(\!a^2\!+\!b^2\!-\!\frac{2b}{R}\right)+\dots
   \right]
\end{array}
\]

{\bf 2.} $\mathcal{E}_c (R)$: Darwin term in the Uehling relativistic correction
\[
\frac{1}{8}\left\langle i \bigl| \Delta U_{vp}(r_1) \bigr| j \right\rangle =
   Z_1\alpha^4 e^{-b R}
   \left[
      \frac{3\pi}{32}\,a
      -\frac{\alpha}{30} \left(3a^2\!+\!b^2\!-\!\frac{2b}{R}\right)
      +\frac{5\pi\,\alpha^2}{384} \, a\left(\!a^2\!+\!b^2\!-\!\frac{2b}{R}\right)+\dots
   \right].
\]

{\bf 3.} $\mathcal{E}_b^{(el)}(R)$: second-order term in the Uehling relativistic correction. In order to evaluate the first two terms of Eq.~(\ref{eb-final}) the following matrix elements are required: \vspace{3mm}
\[
\frac{\alpha^2}{2}\left\langle i \left|\left(\frac{Z_1}{r_1}\right)\,U_{vp}(r_1)\right| j \right\rangle =
   Z_1^2\alpha^4 e^{-b R}
   \left[
      -\frac{3\pi}{16}
      +\frac{2\alpha}{15}\,a
      -\frac{5\pi\,\alpha^2}{1152} \left(3a^2\!+\!b^2\!-\!\frac{2b}{R}\right)+\dots
   \right],
\]
\[
\frac{\alpha^2}{2}\left\langle i \left|\left(\frac{Z_2}{r_2}\right)\,U_{vp}(r_1)\right| j \right\rangle =
   Z_1Z_2\alpha^5 \frac{e^{-b R}}{R}
   \left[
      -\frac{2}{15}
      +\frac{5\pi\alpha}{192}\,a+\dots
   \right].
\]
\[
\begin{array}{@{}l}\displaystyle
\alpha^2\left\langle i \left|\left(-Z_1^2\ln{r_1}\right)\,U_{vp}(r_1)\right| j \right\rangle =
   Z_1^3\alpha^5 e^{-b R}
\\[3mm]\displaystyle\hspace{17mm}\times
   \left[
      \left(
         \frac{4}{15}\ln{\alpha}
         -\frac{4}{15}\gamma_E
         -\frac{17}{225}
      \right)
      +\pi\alpha\left(
         \frac{5}{96}\ln{\alpha}
         -\frac{5}{96}\gamma_E
         -\frac{5\ln{4}}{96}
         +\frac{107}{1152}
      \right)a +\dots
   \right],
\end{array}
\]
\[
\alpha^2\left\langle i \left|\left(-Z_2^2\ln{r_2}\right)\,U_{vp}(r_1)\right| j \right\rangle =
Z_1 Z_2^2 \alpha^5 e^{-b R} \ln (R) \left[ \frac{4}{15} - \frac{5\pi\alpha}{96}\,a +\dots  \right].
\]
We have checked that the numerical results obtained by using these expansions coincide (at the required level of accuracy) with those of numerical integration with the approximate form~\cite{Fullerton76} of the Uehling potential. For example the values of $\mathcal{E}_{vp}(R)$ agree within 8 digits for the whole range of internuclear distances. For illustration we give both values at the equilibrium distance $R = 2.0$~a.u.: $\mathcal{E}_{vp}^{(exp)} (R=2.0) = -0.1108301844 \; \alpha^3$, $\mathcal{E}_{vp}^{(num)}(R=2.0) = -0.1108301853 \; \alpha^3$.

\subsection*{Three-body problem: matrix elements of $U_{vp}(r_1)/r_1$}

For the calculation of $\Delta E_b^{sing}$ the following integral is required (Using the notations of~\cite{Karr13}):
\begin{equation}
I_{-1,1,1}^{(i)} (\alpha,\beta,\gamma) = \int\!\int\!\int dr_1 dr_2 dr_{12} \frac{r_2 r_{12}}{r_1} \; e^{- \alpha r_1 - \beta r_2 - \gamma r_{12}} \; \int_1^{\infty} du \, e^{-2xur_i} \frac{\sqrt{u^2 - 1} \left( 2u^2 + 1 \right)}{u^4}
\end{equation}
where $x = 1/(\alpha_{fsc} m_1)$ ($\alpha_{fsc}$ is the fine-structure constant). Changing the order of integrations we obtain
\begin{equation}
I_{-1,1,1} (\alpha,\beta,\gamma) = \int_1^{\infty} du \, e^{-2xur_i} \frac{\sqrt{u^2 - 1} \left( 2u^2 + 1 \right)}{u^4} \int\!\int\!\int dr_1 dr_2 dr_{12} \frac{r_2 r_{12}}{r_1} \; e^{- \alpha r_1 - \beta r_2 - \gamma r_{12}}
\end{equation}
The integral over spatial coordinates is~\cite{Korobov02}
\begin{eqnarray}
\Gamma_{-1,1,1} (\alpha,\beta,\gamma) &=& \frac{\beta^2 + \gamma^2 + \alpha\beta + \alpha\gamma}{2(\beta-\gamma)^2 (\beta + \gamma)^2 x^2} \; \frac{1}{(u+a)(u+b)} + \frac{\beta + \gamma}{(\beta-\gamma)^2 (\beta + \gamma)^2 x} \; \frac{u}{(u+a)(u+b)} \nonumber \\
&& - \frac{8\beta\gamma}{(\beta-\gamma)^3 (\beta + \gamma)^3} \; \left[ \ln \left( 1+\frac{a}{u} \right) - \ln \left( 1+\frac{b}{u} \right) \right]
\end{eqnarray}
with $a = (\alpha + \beta)/2x$, $b = (\alpha + \gamma)/2x$. Then we get
\begin{equation}
I_{-1,1,1} (\alpha,\beta,\gamma) = \frac{\beta^2 + \gamma^2 + \alpha\beta + \alpha\gamma}{2(\beta-\gamma)^2 (\beta + \gamma)^2 x^2} \; I_1 (a,b) + \frac{\beta + \gamma}{(\beta-\gamma)^2 (\beta + \gamma)^2 x} \; I_4 (a,b) - \frac{8\beta\gamma}{(\beta-\gamma)^3 (\beta + \gamma)^3} \; \left( I_5 (a) - I_5 (b) \right) \; ,
\end{equation}
where
\begin{equation}
I_1(a,b) = \int_1^{\infty} du \frac{\sqrt{u^2-1} \left(2u^2 + 1 \right)}{u^4(u+a)(u+b)}
\end{equation}
was already obtained in~\cite{Karr13}, and
\begin{equation}
I_4(a,b) = \int_1^{\infty} du \frac{\sqrt{u^2-1} \left(2u^2 + 1 \right)}{u^3(u+a)(u+b)} \; ,
\end{equation}
\begin{equation}
I_5(a) = \int_1^{\infty} du \frac{\sqrt{u^2-1} \left(2u^2 + 1 \right) \ln \left( 1 + \frac{a}{u} \right) }{u^4} \; .
\end{equation}
We find, for $a \neq b$,
\begin{equation}
I_4(a,b) = \frac{4ab(a+b) - \pi (2a^2 + 2ab + 2b^2 +3 a^2 b^2)}{4 a^3 b^3} - \frac{\sqrt{1-a^2} (1+2a^2) \arccos (a)}{a^3 (a-b)} + \frac{\sqrt{1-b^2} (1+2b^2) \arccos (b)}{b^3 (a-b)},
\end{equation}
and
\begin{equation}
I_5(a) = \frac{ 12 a + 56 a^3 - \pi (6 + 27 a^2) + 9 \pi^2 a^3 + 12 \sqrt{1-a^2} (1+5a^2) \arccos (a) }{36a^3} - \arccos (a)^2.
\end{equation}
In order to obtain the last expression, it is convenient to calculate $\frac{dI_5}{da}$, and then integrate with respect to $a$.

\subsection*{Three-body problem: matrix elements of $\Delta_{\br_1} U_{vp}(r_1)$}

Using that
\begin{equation}
\Delta\left(\frac{e^{-\Lambda r}}{r}\right) =
   -4\pi\delta(\mathbf{r})
   +\Lambda^2\>\frac{e^{-\Lambda r}}{r},
\end{equation}
and inverting the order of integration as previously, it can be seen that the following integral is required for calculation of $\Delta E_c$:
\begin{equation}
I_c (\alpha,\beta,\gamma) = \int_1^{\infty} du \frac{\sqrt{u^2-1} \left(2u^2 + 1 \right)}{u^4} \; \left( 4 x^2 u^2 I_{0,1,1} (\alpha + 2ux,\beta, \gamma) - \frac{4}{(\beta+\gamma)^3} \right)
\end{equation}
Algebraic manipulations lead to
\begin{equation}
I_c (\alpha,\beta,\gamma) = \frac{1}{x(\beta + \gamma)} \left\{ \frac{1}{2 x} I_6 (a,b) + \frac{1}{\beta + \gamma} I_7(a,b) - \frac{\alpha^2 + \alpha\beta +\alpha\gamma + \beta\gamma }{x(\beta + \gamma)^2} I_1(a,b) - \frac{2(2\alpha + \beta + \gamma )}{(\beta + \gamma)^2} I_4(a,b) \right\}
\end{equation}
with
\begin{equation}
I_6(a,b) = \int_1^{\infty} du \frac{\sqrt{u^2-1} \left(2u^2 + 1 \right)}{u^2(u+a)^2 (u+b)^2} \; ,
\end{equation}
\begin{equation}
I_7(a,b) = \int_1^{\infty} du \frac{\sqrt{u^2-1} \left(2u^2 + 1 \right)}{u^2} \left( \frac{1}{(u+a)^2 (u+b)} + \frac{1}{(u+a) (u+b)^2} \right) \; .
\end{equation}
One obtains
\begin{eqnarray}
I_6 (a,b) &=& -\frac{2 (a^2 + ab + b^2 + 2 a^2 b^2)}{a^2 b^2 (a-b)^2} + \frac{\pi (a+b)}{a^3 b^3} + \frac{1}{(a-b)^3} \left( \frac{ ( 2a^5 + 2a^4 b - a^3 -a^2 b - 4a + 2b ) \arccos (a)}{a^3 \sqrt(1-a^2)} \right. \nonumber \\
&& \left.- \frac{ ( 2b^5 + 2b^4 a - b^3 -b^2 a - 4b + 2a ) \arccos (b)}{b^3 \sqrt(1-b^2)} \right) \; ,
\end{eqnarray}
\begin{equation}
I_7(a,b) = \frac{ -2ab(a+b) + \pi(a^2 + ab + b^2)}{a^3 b^3} + \frac{(2 - a^2 + 2 a^4) \arccos (a)} {a^3 \sqrt(1-a^2) (a-b)} - \frac{(2 - b^2 + 2 b^4) \arccos (b)} {b^3 \sqrt(1-b^2) (a-b)} \; .
\end{equation}

\end{document}